# State-space modeling of dynamic genetic networks

July 16, 2013

**Abstract**

The genomic reality is a highly complex and dynamic system. The recent development of high-throughput technologies has enabled researchers to measure the abundance of many genes (in the order of thousands) simultaneously. The challenge is to unravel from such measurements, gene/protein or gene/gene or protein/protein interactions and key biological features of cellular systems. Our goal is to devise a method for inferring transcriptional or gene regulatory networks from high-throughput data sources such as gene expression microarrays with potentially hidden states, such as unmeasured transcription factors (TFs), which satisfies certain Markov properties. We propose a dynamic state space representation. Our method is based on an EM algorithm with an incorporated Kalman smoothing algorithm in the E-step, a bootstrap for confidence intervals to infer the networks and the AIC for model selection. The state space model is an approach with proven effectiveness to reverse engineer transcriptional networks. The proposed method is applied to time course microarray data obtained from well established T-cell. When we applied the method to the T-cell data, we obtained 4, as the optimum number of hidden states. Our results support interesting biological properties in the family of Jun genes. The following genes were mostly seen as regulatory genes. These genes includes FYB, CCNA2, AKT1, TRAF5, CASP4, and CTNNB1. We found interaction between Jun-B and SMN1, and CDC2 activates Jun-D. We found few significant interactions or one-to-one correspondence among the 4 putative transcription factors. Among the important key genes in terms of outward-directed edges, we found genes such as CCNA2, JUNB, CDC2, CASP4, JUND to have a high degree of connectivity. R Computer source code is made available at our website at http://www.math.rug.nl/stat/Main/Software.

## 1 Introduction

Since the turn of the century a new scientific field has started to emerge: system biology has been brought to the fore front of life-science based research and development, (Bernhard, 2011). It is a biology-based, but inter-disciplinary field that focuses on the systematic study of complex interactions in biological systems. The aim of this holistic approach is to discover new emergent properties that may arise from the systemic view, which would not arise from reductionist approaches. The concept of gene network is central in system biology. We view networks as comprising of nodes (the genes) and the links (chemical reactions) between them. It describes the idea of the stability and interconnectedness of molecular reactions. The challenge is to give this a precise statistical interpretation. In recent times, expression level of many of genes can be measured simultaneously through many techniques including DNA hybridization arrays (Wen et al., 1998; Derisi et al., 1997). A major challenge in system biology is to uncover, from such measurements, gene/protein interactions and key biological features of cellular systems. We present a statistical method that infers the complexity, the dependence structure of the network topology and the functional relationship between the genes; we also deduce the kinetic structure of the network. Our approach is based on the linear Gaussian state space models (SSM) (Fahrmeir and Kunstler, 2009; Fahrmeir and Wagenpfeil, 1997), (Zoubin, 2001; Yamaguchi et al., 2007) applied to real experimental data obtained from a well established model of T-cell activation, where relevant genes are monitored across various time points. Most publications only consider static Bayesian networks (Nir et al., 2000), that model discretized data but incorporate hidden variables. However there has been an increasing need for dynamic modeling that assumes the observed gene expression in the form of mRNA to be continuous time series gene expression data and at the same time incorporate unknown factors such as hidden variables. We build a dynamic model of observed variables (RNA transcripts) and unobserved quantities commonly unmeasured protein regulators, and the relationships between the hidden state variables and the observed RNA transcripts. We infer the model structure as a biological network by estimating model interaction parameters through the EM algorithm (Dempster. et al., 1997; Beal et al., 2005; Ghahramani and Hinton, 1996) combined with the Kalman smoothing algorithm (Shumway and Stoffer, 2005; Meinhold and Singpurwalla, 1983) in the context of maximum likelihood estimation. We use the bootstrap approach in (Efron, 1979) to infer the complex transcriptional response of the network and to reveal interactions between components.

Choosing SSM to model network kinetics has a number of advantages. Most importantly, it allows the inclusion of hidden regulators which can either be unobserved gene expression values or transcription factors (TFs). It can be used to model gene-gene or gene-protein interactions. The hidden variables also allow us to handle noisy continuous measurements which represent the observed gene expression level at each time point. Next, the parameter estimates obtained through the EM algorithm and the state estimates from the Kalman filter have been shown to be consistent and asymptotically normal under some general conditions; (Ljung and Caines, 1979; Dent and Min, 1978). The EM algorithm itself guarantees at least a monotonically increasing likelihood.

Model selection or determining a suitable dimension of the hidden state is an additional complication. Rangel et al. (2004) approached the problem of deciding on a suitable dimension of the hidden state through cross validation. In their approach, they continuously increased the dimension of the hidden states and monitored the predictive likelihood using the test data; one major drawback of this approach is that it is very slow.

Several authors have exploited Kalman filtering and SSM of gene expression and used them to reverse engineer transcriptional networks. To this effect, Fang-Xiang et al.

(2004), in modeling gene regulatory networks, used a two-step approach. In the first step, factor analysis is employed to estimate the state vector and the design matrix; the optimum dimension of the state vector $k$ was determined by minimum BIC. In the second step, the matrix representing protein-protein translation is estimated using least squares regression. Rangel et al. (2004) have applied SSM to T-cell activation data in which a bootstrap procedure was used to derive a classical confidence interval for parameters representing gene-gene interaction through a re-sampling technique. Beal et al. (2005) approached the problem of inferring the model structures of the SSM using variational approximations in the Bayesian context through which a Variational Bayesian treatment provides a novel way to learn model structure and to identify optimal dimensionality of the model. Recently, Bremer and Doerge (2009) used SSM to rank observed genes in gene expression time series experiments according to their degree of regulation in a biological process. Their technique is based on Kalman smoothing and maximum likelihood estimation techniques to derive optimal estimates of the model parameters; however, not much attention was paid to the dimension of the hidden state.

In this chapter, we demonstrate how the EM algorithm with the Kalman smoothing algorithm are used in the maximum likelihood set-up to reverse engineer transcriptional networks from gene expression profiling data. By so doing, we are able to add some useful interpretations to the model. We use the minimum AIC to determine the hidden state's optimal dimension.

The rest of the chapter is organized as follows. In section 2, we introduce the model, and give it a precise mathematical and biological interpretation. Section 3 describes the inference method including the model selection procedure. Identifiability is also discussed briefly and we point out that if we simply estimate parameters of SSM without further constraints on parameter space, the parameters are not identifiable and the EM algorithm may get stuck to a local maximum. Section 4 is the application of our model to real data (T-cell data) through a bootstrap procedure where we identify the network kinetics, by identifying genetic regulatory networks. We summarize our results, analyze their statistical significance and their biological plausibility in section 5, and conclude with a discussion of the method used, possible extension and a summary of related work in section 6.

## 2  Genomic state space model

Linear Gaussian state space models, also known as linear dynamical systems or Kalman filter models (Brown and Hwang, 1997; Dewey and Galas, 2000), are a class of dynamic Bayesian networks that relate temporary observation measurements $y_t$ to some hidden state variable $\theta_t$. We consider a sequence $(y_1, ..., y_T)$ of $p$-dimensional real-valued observation vectors through time, which we shall simply denote by $y_{1:T}$, representing a gene expression data matrix with $p$ rows and $T$ columns, where $p$ and $T$ are the number of genes and the measuring time points, respectively. The model assumes that the evolution of the hidden variables $\theta_t$ is governed by the state dynamics, which follows a first-order Markov process and is further corrupted by a Gaussian intrinsic biological noise $\eta_t$. However, these hidden variable are not directly accessible but rather can be inferred through the observed data vector, $y_t$, namely the quantity of mRNA produced by the gene at time $t$. The observation $y_t$ is a possibly time-dependent linear transformation of a $k$- dimensional real-valued $\theta_t$ with observational Gaussian noise $\xi_t$. The model is given by assuming $n_R$ biological replicates as follows:

$$\begin{cases} \theta_{tr} = F\theta_{t-1,r} + Ay_{t-1,r} + \eta_{tr} \\ y_{tr} = Z\theta_{t,r} + By_{t-1,r} + \xi_{tr} \end{cases} \quad (1)$$

where $r = \{1, 2, ..., n_R\}$, $F$, $A$, $Z$ and $B$ represent the model interactions parameters of dimensions compatible with the matrix operations required in 1. The terms $\eta_t$ and $\xi_t$ are zero-mean independent system noise and measurement noise, respectively with

$$E(\eta_t \eta_t^{'}) = Q, \qquad E(\xi_t \xi_t^{'}) = R \quad (2)$$

Both $Q$ and $R$ are assumed to be diagonal in many practical applications. The initial state $\theta_0$ is independently Gaussian distributed with mean $a_0 = 0$ and covariance $Q_0$. This model is more complex and represents an extension of the standard SSM described in chapter 1 (Fahrmeir and Kunstler, 2009) as it includes various forms of feedback and can also be extended to include additional covariates.

Figure 1: **A 2 gene network representing an input-dependent SSM for Gene regulation with the vector of observed gene expression ($y_t$) and the hidden regulators of gene expression ($\theta_t$) at $3$ different time points, where F, A, Z, and B correspond to the matrices in Equation (1).**

A mathematical representation of the model is depicted in Figure 1 indicating two dynamics, the state and the observed, across 3 consecutive time points, where we assumed $k = p = 2$. The model in Figure 1 assumes RNA-protein translation at two consecutive time points through the matrix $A$, and instantaneous protein-RNA transcription through $Z$. From a biological point of view, the model describes two fundamental stages in gene regulation which are in conformity with the central dogma which states information flows from DNA via RNA to proteins through transcription and translation. The translation matrix $A$, also known as observation-to-state matrix, is of dimension $(k \times p)$, and models the influence or the effects of the gene expression values from previous time steps on the hidden states and $B$ is the $(p \times p)$ matrix indicating the direct gene-gene interactions. The state dynamic matrix $F$ describes the temporal development of the regulators or the evolution of the transcription factors from previous time step $t-1$ to the current time step $t$ and is of dimension $(k \times k)$. It provides key information on the influences of the hidden regulators on each other. The observation dynamics matrix $Z$ relates the transcription factors to the RNAs at a given time point. We now collect the model interaction parameters into a single vector $\varphi$ i.e $\varphi = \{G, Q, R, Q_0\}$ where $G = \begin{bmatrix} B & Z \\ A & F \end{bmatrix}$ represents our genomic graph of interactions.

## 3 Inference

### 3.1 Identifiability issues

Briefly speaking, a parameter of a dynamic system is said to be identifiable given some data if only one value of this parameter can produce the observed likelihood. The identifiability property is important because it guarantees that the model parameter can be determined uniquely from the available data. The poor identifiability of the SSM stems from the fact that given the original model (Equation 1), and with the linear transformation of the state vector $\theta_t^* = T\theta_t$, where $T$ is a non-singular matrix, we can find a different set of parameter vectors

$$\hat{\varphi}^* = \left\{\hat{G}^*, \hat{Q}^*, \hat{R}^*\right\}$$

that give rise to the same observation sequence $\{y_t, t = 1, 2, ..., T\}$ having the same likelihood as the one generated by the parameter vector $\varphi$. Hence, if we place no constraints on $F$, $A$, $Z$, $B$ and possibly $Q$ and $R$, there exists an infinite space of equivalent solutions $\hat{\varphi}$ all with the same likelihood value. To overcome such identifiability issues, further restrictions have to be imposed on the model. In our work, we subject $R$ to be diagonal matrix, $Q$ to be identity matrix and $Q_0$ to be a fixed diagonal matrix. Subjecting $Q$ to be identity only affects the scale of $\theta$ and matrices $A$ and $Z$. The matrices $A$ and $Z$ are then identifiable from the data, which can be seen from the marginal covariance matrix of $y$, $\Sigma_y$. The latter, according to Schur complement Horn and Johnson (1990) is given by

$$\Sigma_y = \left(K_{yy} - K_{y\theta} K_{\theta\theta}^{-1} K_{\theta y}\right)^{-1} \tag{3}$$

where $K_{yy}$ specifies the concentration matrix of the conditional statistics of the observed variables given the hidden variables and is usually sparse and the quantity $K_{y\theta} K_{\theta\theta}^{-1} K_{\theta y}$ is of low rank; (Chandrasekaran et al., 2010).

We further assume that the errors $\{\eta_t, t = 1, ..., T\}$ and $\{\xi_t, t = 1, ..., T\}$ are jointly normal and uncorrelated. Also the number of time points or biological observations in microarray data are typically much smaller than the number of genes. This fundamental problem of high-dimensional statistical modeling of micro array data demands some care in the estimation of model parameters in the state space model. This problem is avoided by making sure that the number of observations exceed the total number of parameters to be estimated

$$pTn_R > p^2 + 2kp + k^2. \tag{4}$$

This further puts the following bound on the dimension of the hidden states as given in Equation 5

$$0 \leq k < -p + \sqrt{pTn_R}. \tag{5}$$

## 3.2 The likelihood function

We now restrict the model interaction parameters into the single vector $\varphi = \{F, A, Z, B, R\}$. As can be seen from Figure 1, the observations at time $t$, $y_{tr}$ are conditioned on the past observations, $y_{(t-1)r}$ and on the regulators $\theta_{tr}$ and also to infer for instance $\theta_{tr}$, we need $\theta_{(t-1)r}$ and $y_{(t-1)r}$. To that effect, under the Gaussian assumption we have the following:

$$\begin{aligned}
\theta_{0r} &\sim \psi_k(\theta_{0r}|0, Q_0) \\
\theta_{tr}|\theta_{(t-1)r}, y_{(t-1)r} &\sim \psi_k(\theta_{tr}|\tilde{\theta}_t, Q) \\
y_{tr}|\theta_t, y_{(t-1)r} &\sim \psi_p(y_{tr}|\tilde{y}_t, R).
\end{aligned}$$

where

$$\tilde{\theta}_t = F\theta_{t-1} + Ay_{t-1},$$
$$\tilde{y}_t = Z\theta_t + By_{t-1},$$

and $\psi(.|\mu, \Sigma)$ is the normal density with mean $\mu$ and variance $\Sigma$.

We now write the marginal likelihood function $l_y^m(\varphi)$ of the data $y$. This is given by

$$\begin{aligned}
l_y^m(\varphi) &= \int \prod_{t=1}^T P(\theta_t|F, A, \theta_{t-1}, y_{t-1}) \times \\
&\quad P(y_t|B, Z, \theta_t, y_{t-1})d\theta \\
&= \int \prod_{t=1}^T \psi(\theta_t|\tilde{\theta}_t, \sigma_\eta^2 I)\psi(y_t|\tilde{y}_t, \sigma_\xi^2 I)d\theta. \tag{6}
\end{aligned}$$

The full log-likelihood function of the complete data $(y_{tr}, \theta_{tr})$ denoted by $l_{y,\theta}(\varphi)$ is for simplicity given by

$$l_{y,\theta}(F, A, Z, B) = \sum_{r=1}^{n_R} l^r_{y_r\theta_r}(F, A, Z, B) \qquad (7)$$

where $l^r_{y_r\theta_r}(F, A, Z, B)$ is the complete log-likelihood of the $r^{th}$ replicate and is given by

$$\begin{aligned}
l^r_{y_r\theta_r}(F, A, Z, B) &= \sum_{t=1}^{T} l_{y_t|\theta_t, y_{(t-1)}}(Z, B) + \sum_{t=1}^{T} l_{\theta_t|\theta_{(t-1)}, y_{(t-1)}}(F, A) \\
&= -\frac{1}{2\sigma_\xi^2} \sum_{t=1}^{T} (y_t - \tilde{y}_t)' (y_t - \tilde{y}_t) - \frac{T}{2}\log(\sigma_\xi^2) \\
&\quad - \frac{1}{2\sigma_\eta^2} \sum_{t=1}^{T} \left(\theta_t - \tilde{\theta}_t\right)' \left(\theta_t - \tilde{\theta}_t\right) - \frac{T-1}{2}\log(\sigma_\eta^2)
\end{aligned} \qquad (8)$$

ignoring constant term.

### 3.3 Joint parameter estimation via EM algorithm

Our aim is to estimate the model parameter $\varphi$ which (excluding $R$) indicates connectivity matrix of the directed genomic graph that maximizes the marginal likelihood function $l_y^m(\varphi)$ given in Equation 6. The integral in Equation 6 is intractable because of the presence of the hidden variables $\theta$. For that matter we use the EM algorithm to learn the parameters of the model. The idea stems from the fact that if we did have the complete data $(y_t, \theta_t)$ it will be straight forward to obtain MLEs of $\varphi$ using multivariate normal theory. In this case, we do not have the complete data and the EM algorithm gives us an iterative method for finding the MLE of $\varphi$ using the observed data $y_t$, by successively maximizing the conditional expectation of the complete data likelihood given the observed values. It is only when we are able to estimate the parameter $\varphi$ that we can expect to obtain some useful interpretation of the biological system network; for example magnitude of the effect of proteins on RNA (part of transcription process) and also estimate networks between the proteins by investigating $F$.

The EM-algorithm for SSM was formulated by Shumway and Stoffer (1982) and Shumway (2000). To this effect the algorithm requires the computation of the conditional expectation of the log-likelihood given the complete data. The algorithm is a two-stage procedure in which we begin with a set of trial initial values for the model parameter to calculate the Kalman smoother. The Kalman smoother is then input into the M-step to update parameter estimates. The algorithm alternates recursively between an expectation step followed by a maximization step.

#### 3.3.1 The expected log-likelihood function: The E-step

This step of the EM algorithm involves the calculation of the first two moments $\theta_t$ of the hidden states. Let **Q** denote the expected log-likelihood. Then from Equation 7, **Q**

becomes

$$
\begin{aligned}
\mathbf{Q}(\varphi|\varphi^*) &= E_\theta\left[l_{y,\theta}(\varphi)|y,\varphi^*\right] \\
&= \sum_{r=1}^{n_R} E_\theta\left[l^r_{y_r,\theta_r}(\varphi)|\varphi^*,y\right] \\
&= \sum_{r=1}^{n_R} E_\theta\left[l^r_{y_r,\theta_r}(Z,B)|y,\varphi^*\right] + \sum_{r=1}^{n_R} E_\theta\left[l^r_{y_r,\theta_r}(F,A)|\varphi^*,y\right]. \\
&= \mathbf{Q}(Z,B) + \mathbf{Q}(A,F).
\end{aligned}
\tag{9}
$$

The calculation of $\mathbf{Q}(\varphi|\varphi^*)$ in Equation 9 involves finding $E(\theta)$ and $E(\theta'\theta)$ (See appendix 8) for each replicate $r$; these forms are supplied by the Kalman smoothing algorithm. The above implies that for each replicate we run the Kalman smoothing algorithm to find the expected hidden states and their variance-covariance components and these are joined together to get $\mathbf{Q}(\varphi|\varphi^*)$.

### 3.3.2 The update equations: The M-step.

A new parameter set $\varphi^{i+1}$ is computed by estimating the parameters that maximize equation 9; the expected log-likelihood function that is

$$\varphi_{next} = \underset{\varphi}{\mathrm{argmax}}\left\{\mathbf{Q}(\varphi|\varphi^*)\right\} \tag{10}$$

These can be solved in closed form in the following manner.

$$\frac{\partial}{\partial \varphi}\mathbf{Q} = 0$$

and then solve for the parameter value that sets the partial derivative to zero. It is also important to note that the partial derivatives are taken with respect to matrices $F$, $A$, $Z$ and $B$. For more details about the derivations, see appendix 8.

The entire EM algorithm can be regarded as alternating between Kalman filtering and smoothing recursions and the normal maximum likelihood estimators as given in the update equations.

### 3.4 Choice of hidden state dimension: $AIC_c$

Model selection or the determination of the optimum dimension of the hidden state $k$ is important to the application of SSM to network reconstruction. Popular criteria for model selection include Akaike's Information Criterion (AIC) (Akaike, 1974) and the Bayesian Information Criterion (BIC) (Schwarz, 1978). Given the log-likelihood function $l$, AIC for a model with $k$-dimensional state vector is given by:

$$AIC(k) = -2l(y_t|\hat{\varphi}_k) + 2P \tag{11}$$

with $P$ the number of estimated parameters, and $l(y_t|\hat{\varphi}^k)$ the log-likelihood of the observed data. As recommended by Burnham and Anderson (2002), we have applied $AIC_c$ for our model selection procedure. The $AIC_c$ is given by

$$AIC_c(k) = -2l(y_t) + 2P\left[\frac{N}{N-P-1}\right] \quad (12)$$

where $N = pTn_R$ represents total number of observations and $P = p^2 + 2kp + k^2$ is the total number of estimated parameters and we settle on the hidden state dimension that has the minimum $AIC_c$, i.e we find k such that

$$k = \underset{k}{\mathrm{argmin}}\left\{AIC_c(k)\right\}. \quad (13)$$

In this case, we successively increase the number of hidden states and monitor the behavior of AICc i.e., for each run of the EM algorithm, we increase $k$.

### 3.5 Network Reconstruction by Bootstrapping

In our procedure, we use a bootstrap approach to find confidence intervals for the parameters defining our model. By so doing we compute the bootstrap distribution of the estimator of $\varphi$.

Let $\hat{\varphi}$ denote the MLE of the parameters defining our model; $\hat{\varphi}$ are estimated using the EM algorithm described in previous section. The essence of the bootstrap procedure is to resample with replacement the data or (the replicates within the original data), and given each new data we can estimate among other things, the bootstrap set of parameters $\left\{\hat{\varphi}_b^*; b = 1, ..., N_b\right\}$ through the EM algorithm. Stated differently for each bootstrap data, the parameters that maximize the likelihood of the bootstrap data are found, and then obtain the sampling distributions of the estimators of the elements of $\varphi$. The results of the bootstrapping are the distribution of the parameters and we proceed to make statistical inferences about those underlying parameters by computing confidence interval for each of them; (Wild et al., 2004; Shumway and Stoffer, 2005)

## 4 Simulation studies

In order to evaluate the performance of our method for analyzing gene expression data, we simulate artificial data and applied our proposed method to the simulated data according to the model described in equation 1 with 10 time points, 50 replicates, $p = 2$ as number of genes, and $k = 2$ TFs. Parameters were initialized as follows: $Z$ and $F$ are assumed to be identity matrices whiles we initialize $A$ to be zero. For $B$ we perform a simple linear regression where we regress current genes on its previous one and $R$ assume the usual variance estimate from the regression. $Q$ and $Q_0$ were fixed and assumed diagonal. We applied the bootstrap procedure to the data and identified the significant and non-significant parameters defining our model or identifying the dynamics of the network. We achieved this by computing bootstrap confidence intervals on element $\varphi_l$ of $\varphi$; it is clear that the confidence intervals will enable us to decide which elements $\varphi_l$ will be set to zero and which will not. The analysis now turn to a decision problem where we formulate two hypotheses, namely,

**Figure 2: Confidence interval of the vectorized form of elements $G_l$ of $G$**

$$H_0 : \varphi_l = 0$$

$$H_1 : \varphi_l \neq 0$$

where rejecting $H_0$ indicates the presence of connection among the genes, meaning that the particular interaction in the matrix is considered to be statistically significant. With $k$ equals 2, we obtained the upper and lower bounds of the confidence interval of the vectorized elements of the bootstrap estimated parameters, in the order of $F_{ij}$, $A_{ij}$, $Z_{ij}$, $B_{ij}$. Figure 2 indicates the upper (red solid line) and lower (blue solid line) bounds calculated based on 99 % confidence interval. The points on the graph are the upper and lower bounds confidence interval of the vectorized elements of the bootstrap estimated parameters, in the order of $F_{ij}$, $A_{ij}$, $z_{ij}$ and $B_{ij}$.

To evaluate the performance of the proposed method for gene regulatory network, we calculate the true positive rate (TPR), false positive rate (FPR) and the $F_1$ score of the matrix $G$ representing the entire genomic interaction. Table 1 shows the simulation result at varying number of replicates. According to Table 1, as the number of replicates $n_R$ increases from 20 to 100, we experience the lowest FPR and the highest TPR at replicate 100. Also the highest $F_1$ score occurs at replicate 100.

| $n_R$ | 20 | 50 | 70 | 90 | 100 |
|---|---|---|---|---|---|
| TPR | 0.50 | 0.40 | 0.57 | 0.42 | 0.75 |
| FPR | 0.14 | 0.36 | 0.22 | 0.33 | 0.16 |
| F-score | 0.40 | 0.36 | 0.615 | 0.46 | 0.66 |

**Table 1: Simulation result for TPR and FPR as the number of replicates $n_R$ increases from 20 to 100**

## 5 Application

**Figure 3: The behavior of AIC as a function of the dimension of state space, k**

For this study, to demonstrate the application of our reverse engineering method, we used publicly available data, the results of two experiments used to investigate the expression response of human T-cells to PMA and ionomicin treatment. The data is a combination of two data set namely tcell.34 and tcell.10. The first data set (tcell.34) contains the temporal expression levels of 58 genes for 10 unequally spaced time points. At each time point there are 34 separate measurements. The second data set (tcell.10) comes from a related experiment considering the same genes and identical time points, and contains 10 further measurements per time point. At each time point there are 44 separate measurements or replicates. It was assumed that the 44 replicates have a similar underlying distribution. See Rangel et al. (2004) for more details. Given that the T-cell is a time course gene expression data with technical replicates we expect more reliable estimation and inference results by applying our method. Corresponding to each gene expression $y_{tr}$, we also generated technical replicates for the hidden variables $\theta_{tr}$. With $p = 58$ genes, $R = 44$ as replicates and $T = 10$, the constraint represented by Equation 4 is satisfied, indicating that we have enough data to estimate our parameters. In total we estimate 3844 parameters; the dimension of the hidden was determined by using AICc as explained in section 3.4. Table 2 shows the behavior of AICc with corresponding $k$'s and Figure 3 shows AICc with some selected values of $k$. It turns out that $k = 4$ is the optimum number of the hidden states as compared to Rangel et al. (2004) and Beal et al. (2005) who obtained 9, 14 respectively under different criteria.

| k    | 2    | 4    | 6    | 8    | 10   | 12   | 14   |
|------|------|------|------|------|------|------|------|
| AICc | 6884 | 3224 | 6110 | 7427 | 7206 | 6939 | 4820 |

Table 2: Estimates of $AIC_c$ as a function of $k$

In essence, we treated the data as a time series measurement data $y_{t_r}$, $t = 1, 2, ..., 10$ and $r = 1, 2, ..., 44$. For each replicate, $y_t$ and $\theta_t$ consist of 58 genes and 4 transcriptions factors respectively, each, measured at 10 different time points, i.e for each replicate $r$, $y_t$ and $\theta_t$ are of dimension (58 $by$ 10), (4 $by$ 10) respectively. Some of these genes include RB1, CCNG1, TRAF5, CLU.... The parameters $Q$ and $Q_0$ were fixed.

**Figure 4: EM algorithm on the T-cell data showing the expression level of the TFs across time**

We then applied the EM algorithm to the data and Figure 4 shows the estimated values of the hidden variables $\theta$ i.e the expression pattern of the 4 transcription factors across time.

Based on the test, with 95% confidence level, we plot the connectivity matrix of the directed genomic network $G$. The output is a directed graph showing connections from one gene expression variable at a given time point $t$ to another gene expression variable whose expression it influences at the next time point, $t + 1$. The arrows indicates the direction of the regulation. The entire directed graph $\hat{G}$ gives 704 genomic interactions. Figure 5 represents a portion of the interaction nertwork $\hat{\varphi}$ where we indicate genes that have at least 2 outwards connections. These genes include CCNA2, JUNB,

CDC2, CASP4, JUND to mention but fews. Figure 6 is the sub-network produced at 95% confidence level and it represents the interaction between, two Jun proteins family namely Jun B and Jun D on one hand, and on other hand, various genes involved in programmed cell death. We also recover the topology of the genes FYB through Figure 7. The structure of the network is visualized using the R package for Network analysis and visualization igraph.

**Figure 5: Subnetwork found representing the genomic interactions $\hat{G}$, of genes with at least 2 outwards connections, node refers to gene expression in the form of proteins or RNAs; empty nodes refer to TFs.**

## 6  Results

Our method has resulted in a relatively fairly sparse networks, especially at 95% confidence interval in accordance with our biological intuition. According to our method, the following genes were mostly seen as regulatory genes. These genes includes FYB, CCNA2, AKT1, TRAF5, CASP4, and CTNNB1. The TFs were also seen to regulate the expression level of most genes as could be seen in Figure 5. The latter displays the interaction within a family of genes with more than 2 outwards directed connections. Our approach has revealed interesting features in the family of Jun genes. The network in Figure 6 provides support for interesting biological properties some of which also confirmed in (Rangel et al., 2004) and (Beal et al., 2005); but we also found new connections. In our work, we found no interaction between the proto-oncogene JUNB and the apoptosis-related cysteine protease gene CASP4 but found interaction between JUNB and MAP3K8. This interaction was also was recovered by Beal et al. (2005). Also Figure 6 reveals that the proto-oncogene JUND represses the expression level of the apoptosis-related cysteine protease gene CASP7 and the cell division cycle 2 (CDC2). This further supports the anti-proliferative and anti-apoptotic role of JUND. Furthermore, in our model, the survival of motor neuron 1 gene SMN1 and CDC2 influence negatively the expression level of JUNB and CASP7 respectively. JUNB also activates the expression level of CDC2 whiles CASP8 also activates the expression level of CDC2. A critical comparison of our Figure 6 to that of similar subnetworks found in the work of Andrea et al. (2010) and Beal et al. (2005) shows that in all the 3 subnetworks, gene CDC2 regulates the expression level of JUND. JUNB activates Caspase-8 in the subnetwork found by Beal et al. (2005) and indirectly regulates Caspase-8 through Caspase-4 in the subnetwork found by Andrea et al. (2010). We found interaction in the form of inhibition between CASP4 and CASP7. However such interaction was not identified by Andrea et al. (2010) but was indirectly recovered in the work of Beal et al. (2005) though gene JUND.

The gene FYN-binding protein gene FYB found to occupy one of the most crucial positions in the network recovered by Rangel et al. (2004) also has a high degree of connectivity in our work; and Figure 7 reveals some crucial genes that are found to be directly connected to FYB. Most importantly, in our model, FYB influences the expression level of genes such as the cell division cycle 2 (CDC2), the proto-oncogene JUND, SOD1. FYB is also seen to be connected to genes such as ID3, MAPK9, API1. Clearly, these results support the fact that FYB mRNA levels are predictive of the expression level of a number of genes. The hidden state dimensionality was found to be 4, a result similar to the work of Andrea et al. (2010) in which they developed an iterative empirical Bayesian procedure with a Kalman filter to estimate the posterior distributions of network parameters. Rangel et al. (2004) found the dimension of the hidden state to be 9 through cross validation technique while Beal et al. (2005) through a variational Bayesian approach obtained 14. At 95% confidence level, we found no significant interactions among the hidden variables or transcription factors. However their role in the transcription process cna not be ignored as the inferred matrix $Z$ representing instantaneous protein-RNA transcription was not sparse signifying that the transcription factors regulate the expression level of most mRNAs.

**Figure 6: Subnetwork found representing the interactions between Jun proteins family and other genes**

**Figure 7: Subnetwork found representing the topology of gene FYB in connection with some selected genes**

## 7  Conclusion

In this chapter, we have developed a novel state space model and applied it to the T-cell data. We used the EM algorithm and the bootstrap technique to infer the structure of gene expression data and the underlying genomic network. Some of the substantial results are similar to those in previous literature, but we also discover new interactions in the form of SMN1 and JUNB on one hand, and CDC2 and Caspase-7 on the other hand. In our model, through Figure 6, the gene SMN1 inhibits the expression of Jun-B. The hidden state dimensionality $k$ was found to be 4, similar to the result in Andrea et al. (2010). The proposed method offers significant advantages over other methods that have recently appeared in the literature. For example, Beal et al. (2005) used a variational Bayesian methodology which is an approximation of the posterior distribution of the parameters, while we did exact inference of the parameters. Rangel et al. (2004) used cross validation as model selection technique which is quite slow as compared to AIC. Bremer and Doerge (2009) used an *ad hoc* method for selecting the hidden state dimensionality $k$, while our method uses a data-driven approach. Also our model allows for dynamic correlation over time, as each observation and hidden state depend explicitly on some function of previous observations as opposed to the model described by Yamaguchi and Higuchi (2006); Perrin et al. (2003); Fang-Xiang et al. (2004). Their model does not allow for RNA-protein translation (the matrix $A$ in our model) and RNA-RNA interactions (the matrix $B$ in our model).

One fundamental assumption in our proposed model is the first-order linear dynamics in the state and observation equations of the SSM. This assumption can only be an approximation to the true nature of a complex biological system since more realistic models of gene regulatory interactions surely include complex interactions or nonlinear relationships. Our linear dynamics assumption is a stepping stone upon which a future model with non-linear dynamics will be explored. We also discovered new interactions that do not find support in the current literature; as as part of our future work we will investigate these interactions further and possibly redefine our model. Furthermore gene regulation tends to be sparse and regularization is fundamental to high-dimensional statistical modeling. In future work we plan to employ a penalized maximum likelihood strategy in the context of the EM algorithm in the state space model .

# 8 Appendix

We outline the derivations of the update estimates in the maximization step especially for matrix $Z$ and $B$, the rest follows in a similar manner.

The update estimates for matrix Z and B.

First we derive the update equation for matrix $Z$. Take the derivative of $H$ with respect to $Z$ and $B$ and equating them to 0. We write $\mathbf{Q}(Z,B)$ as

$$\mathbf{Q}(Z,B) = \sum_{r=1}^{n_R} E_{\theta,\varphi^*} \left[ l^r_{y_r,\theta_r}(Z,B) \right] \tag{14}$$

then

$$\begin{aligned}
\mathbf{Q}(Z,B) &= -\sum_{r=1}^{n_R}\sum_{t=1}^{T} y'_{tr} y_{tr} + 2\sum_{r=1}^{n_R}\sum_{t} E(\theta'_{tr} Z y_{tr}) \\
&+ 2\sum_{r=1}^{n_R}\sum_{t} y'_{(t-1)r} B' y_{tr} - \sum_{r=1}^{n_R}\sum_{t}^{T} Z E(\theta'_{tr}\theta_{tr} Z') \\
&- 2\sum_{r=1}^{n_R}\sum_{t}^{T} E(\theta'_{tr} Z' B y_{(t-1)r}) - \sum_{r=1}^{n_R}\sum_{t}^{T} B' y'_{(t-1)r} y_{(t-1)r} B
\end{aligned}$$

Setting $\frac{\partial}{\partial Z}\mathbf{Q}(Z,B)$ and $\frac{\partial}{\partial B}\mathbf{Q}(Z,B)$ equal 0 result in two linear system of equations in the form:

$$0 = -\frac{1}{2\sigma_{\xi_{tr}}^2}\sum_{r=1}^{n_R}\sum_{t=1}^{T}[-2y_{tr}E(\theta'_{tr})$$
$$+2ZE(\theta_{tr}\theta'_{tr}) + 2By_{(t-1)r}E(\theta'_{tr})] \tag{15}$$

and

$$0 = -\frac{1}{2\sigma_{\xi_{tr}}^2}\sum_{r=1}^{n_R}\sum_{t=1}^{T}[-2y_{(t-1)r}y'_{tr} + 2y_{(t-1)r}E(\theta'_{tr})Z'$$
$$+2(y_{(t-1)r}y'_{(t-1)r}B')] \tag{16}$$

Equations 15 and 16 could also be re-written as

$$\hat{Z}\hat{\theta}\hat{\theta}' = y\hat{\theta}' - \hat{B}L(y)\hat{\theta}' \tag{17}$$

$$\hat{B}L(y)L(y)' = yL(y)' - \hat{Z}\hat{\theta}L(y)' \tag{18}$$

where $L(y)$ in Equations 17 and 18 is the shift operator on matrix $y$, and

$$\hat{\theta} = \sum_{r=1}^{n_R}\sum_{t=1}^{T}E(\theta_{tr}) \tag{19}$$

$$\hat{\theta}\hat{\theta}' = \sum_{r=1}^{n_R}\sum_{t=1}^{T}E(\theta_{tr}\theta'_{tr}) \tag{20}$$

Equations 17 further implies

$$\hat{Z} = \left(y\hat{\theta}' - \hat{B}L(y)\hat{\theta}'\right)\left(\hat{\theta}\hat{\theta}'\right)^{-1} \tag{21}$$

Now we substitute equation 21 into equation 18 giving rise to

$$\begin{aligned}\hat{B}L(y)L(y)' &= yL(y)' - \left(\left(y\hat{\theta}' - \hat{B}L(y)\hat{\theta}'\right)\left(\hat{\theta}\hat{\theta}'\right)^{-1}\right)\hat{\theta}L(y)' \\ &= yL(y)' - y\hat{\theta}'\left(\hat{\theta}\hat{\theta}'\right)^{-1}\hat{\theta}L(y)' \\ &\quad + \hat{B}L(y)\hat{\theta}'\left(\hat{\theta}\hat{\theta}'\right)^{-1}\hat{\theta}L(y)'\end{aligned} \tag{22}$$

Rearranging Equation 22, we have

$$\hat{B}\left[L(y)L(y)' - L(y)\hat{\theta}'\left(\hat{\theta}\hat{\theta}'\right)^{-1}\hat{\theta}L(y)'\right] = \\ \left[yL(y)' - \hat{\theta}'\left(\hat{\theta}\hat{\theta}'\right)^{-1}\hat{\theta}L(y)'\right] \quad (23)$$

Finally we have

$$\hat{B} = yRL(y)'\left[L(y)RL(y)'\right]^{-1} \quad (24)$$

where

$$R = I - \hat{\theta}'\left(\hat{\theta}\hat{\theta}'\right)^{-1}\theta \quad (25)$$

At this stage we substitute 24 in to 21 to solve for matrix $Z$ and we get the following:

$$\hat{Z} = \left[y\hat{\theta}' - yRL(y)'\left[L(y)RL(y)'\right]^{-1}L(y)\hat{\theta}'\right]\left(\hat{\theta}\hat{\theta}'\right)^{-1} \\ = y\left[I - RL(y)'\left[L(y)RL(y)'\right]^{-1}L(y)\right]\hat{\theta}'\left(\hat{\theta}\hat{\theta}'\right)^{-1} \quad (26)$$

Equations 26 and 24 are the update equations in the maximization step used to infer the parameters in the observation dynamics. In the same manner we derive the updates equations for $A$ and $F$ for the model interaction parameters in the state dynamics model.